# The Postulate of the Three Regimes of Economic Growth Contradicted by Data

Ron W Nielsen[1]

Environmental Futures Research Institute, Gold Coast Campus, Griffith University, Qld, 4222, Australia

Economic growth in Western Europe, Eastern Europe, Asia, countries of the former USSR, Africa and Latin America were analysed. It is demonstrated that the fundamental postulate of the Unified Growth Theory about the existence of the three regimes of growth (Malthusian regime, post-Malthusian regime and sustained-growth regime) is contradicted by data. These regimes did not exist. In particular, there was no escape from the Malthusian trap because there was no trap. Economic growth in all these regions was not stagnant but hyperbolic. Unified Growth Theory is fundamentally incorrect. However, this theory is also dangerously misleading because it claims a transition from the endless epoch of stagnation to the new era of sustained economic growth, the interpretation creating the sense of security and a promise of prosperity. The data show that the opposite is true. Economic growth in the past was sustained and secure. Now, it is supported by the increasing ecological deficit. The long-term sustained and secure economic growth has yet to be created. It did not happen automatically, as suggested incorrectly by the Unified Growth Theory.

**Introduction**

There is no science without data but there is also no science without scientific analysis of data. We can have excellent data but if we do not analyse them properly we are likely to draw incorrect conclusions. A perfect example is the Unified Growth Theory (Galor, 2005a, 2011). Excellent data (Maddison, 2001) were available and even used during its formulation but they were never properly analysed. Now, it can be easily demonstrated that the fundamental postulates of this theory are repeatedly contradicted by data, making it fundamentally incorrect and, consequently, unacceptable.

Many attractive theories and explanations can be formulated but if they are not based firmly on the rigorous analysis of data they are only, at best, just interesting stories. They may contain elements of truth but folklores of many cultures are full of such stories and they also contain elements of truth. Fantasy and leaps of faith might be inspiring and productive even in scientific research but they have to be soon tested by the scientific process of investigation.

---

[1]AKA Jan Nurzynski, r.nielsen@griffith.edu.au; ronwnielsen@gmail.com; http://home.iprimus.com.au/nielsens/ronnielsen.html

However, if one leap of faith is followed by another, if one fantasy creates another, then we no longer deal with science but with fiction. It is then easy to loose scientific perspective and defend emotionally the widely-accepted dogmas, based on faith.

Any theory that cannot be checked by data is unscientific even if it is based on scientifically attractive ideas. Such a theory has to be put aside until it can be checked by relevant data. Even if a theory is confirmed by many sets of data it can be still challenged by a single set of contradicting data. Any theory contradicted by just one set of good data has to be either revised or rejected. Any research, any intellectual activity, which ignores these fundamental principles of scientific investigation is unscientific even if it is intellectually stimulating and attractive.

In science it is important to look for data confirming theoretical explanations but it is even more important to discover contradicting evidence, because data confirming a theory confirm only what we already know but contradicting evidence may lead to new discoveries.

If scientific analysis of data is found to be in agreement with a proposed theory, this theory may then be considered to be supported by data and its explanations of studied phenomena may be then accepted. However, if just one set of data is found to be in contradiction with this theory, then this theory can no longer be accepted in its original form. It has to be then either modified to bring it in agreement with data, or rejected if such modification is impossible. There is no scientific gain in accepting such a theory. On the contrary, its continuing acceptance is detrimental to science.

When an incorrect theory is rejected we can then look for a better explanation of studied phenomena. There are no sentimental values in scientific research and no emotional attachments, and any scientist should be prepared to have his or her theories challenged by science.

**Unified Growth Theory**

Currently, the most complete theory of the historical economic growth is the Unified Growth Theory (Galor, 2005a, 2011). It follows closely the traditional interpretations of economic growth. One of its fundamental postulates is the existence of the three regimes of growth. It claims that the historical economic growth in various countries and regions can be divided into *three distinctly different regimes of growth governed by distinctly different mechanisms*. We shall show that these three regimes did not exist.

The alleged regimes are:

1. The regime of Malthusian stagnation. According to Galor, and indeed according to the currently accepted interpretations, this regime "characterized most of human history" (Galor, 2005a, p. 178). Economic growth was allegedly in the endless state of stagnation described as the Malthusian trap or "the Malthusian steady-state equilibrium" (e.g. Galor, 2005a. pp. 236, 237, 244). Galor claims that this epoch of stagnation commenced in 100,000 BC (Galor 2008a, 2012a) and was terminated in around AD 1750, or around the time of the Industrial Revolution, 1760-1840 (Floud & McCloskey, 1994), in developed regions and around AD 1900 in less-developed regions.

    The beginning of this regime in 100,000 BC is highly speculative because Maddison's data do not extend to the BC era. Furthermore, the emergence of *Homo Sapiens* is usually claimed to have been around 200,000 BC or maybe even earlier (Weaver, Roseman & Stringer, 2008). We simply do not know about the economic growth in

such a distant past because we do not have relevant data. Judging by the available evidence (Nielsen, 2016a, 2016b, 2016c), the growth was probably hyperbolic but whatever we might want to suggest will be based on speculations. However, we do not have to go so far back in time to test the Unified Growth Theory because the postulate of the existence of the three regimes of growth cannot be even tested using the economic growth data for the BC era. Even if such data were available they would be inapplicable for this purpose because the existence of the three regimes of growth is not claimed for the BC era but only for the AD era. The data we need to use are the data of Maddison (2001, 2010) because they cover the time when the alleged three regimes were supposed to have existed.

2. The post-Malthusian regime. According to Galor (2008a, 2012a), this regime was between AD 1750 and 1870 for developed regions but it commenced a little later, in around AD 1900, for less-developed regions. Thus, the alleged escape from the Malthusian trap and the commencement of the fast economic growth occurred around the onset of the Industrial Revolution for developed regions and a little later for less-developed regions.

3. The sustained-growth regime. According to Galor (2008a, 2012a), this regime commenced around AD 1870 for developed regions.

The general idea of this interpretation of the historical economic growth is that after the endless epoch of "the Malthusian steady-state equilibrium," humans were finally able to break through the impenetrable barrier of stagnation, escape the Malthusian trap and enter into a new era of sustained and rapid economic growth. This is not only incorrect but also dangerous concept because the data describing the historical economic growth (Maddison, 2001, 2010) present a diametrically opposite interpretation. The economic growth was sustained and secure in the past (Nielsen, 2016a) but now it entered a stage of the insecure future (Nielsen, 2015a).

We shall now demonstrate that *Golor's concept of the three regimes of growth is contradicted by the economic growth data* (Maddison, 2010). We shall show that his three regimes of growth have no correlation with data and no positive connection with the real world. Within the range of the mathematically-analysable data, there was no stagnation and no transition to a fast economic growth, described as the sustained-growth regime or the modern-growth regime. We shall show that during this alleged new, fast-increasing and sustained-growth regime, economic growth started to be diverted from the fast-increasing historical hyperbolic trajectories to *slower* trajectories.

Historical economic growth, global and regional, was so well sustained that it followed stable hyperbolic trajectories. However, such trajectories escape to infinity at a fixed time and any growth, which follows them, has to be, at a certain stage, diverted to a slower trajectory. Economic growth, global and regional, is now diverted to slower trajectories. However, the momentum gained during the sustained historical growth keeps on propelling the economic growth along trajectories, which are still increasing too fast to feel comfortable about their future.

Galor's Unified Growth Theory is *not* based on the scientific analysis of data. He had access to the excellent set of data (Maddison, 2001) but he did not analyse them. Now, precisely the same data can be used to show that his theory is fundamentally incorrect.

Regrettably, Unified Growth Theory is based on impressions created by the customary disfigured presentation of data (Ashraf, 2009; Galor, 2005a, 2005b, 2007, 2008a, 2008b, 2008c, 2010, 2011, 2012a, 2012b, 2012c; Galor and Moav, 2002; Snowdon & Galor, 2008).

Example of such distorted presentation of data is shown in Figure 1. This way of handling data is a perfect prescription for drawing incorrect conclusions.

In science, data are treated with respect because the primary aim of science is to discover the truth, and for this purpose there is nothing as reliable as good sets of data. Not all data can be accepted but we have to have good reasons for rejecting data. If reasons for rejecting data are unacceptable, then reasons for rejecting data have to be rejected.

Many attractive theories and explanations may be formulated but they all have to pass the test of data. Without such a test, they are just stories, which might or might not be true.

Galor's predecessors might be excused for believing in the existence of Malthusian stagnation and in the dramatic impact of the Industrial Revolution on changing the economic growth trajectories because they were using strongly limited information. They had no access to the excellent source of data published by the world-renown economist (Maddison, 2001). Galor not only had access to these data but he also used them repeatedly during the formulation of his theory but unfortunately he distorted them so much that they were creating an impression of being in agreement with his postulates.

In our discussion we shall use the latest data describing economic growth (Maddison, 2010). This publication contains some additional information but any of Maddison's compilations, the compilation used by Galor or this new compilation, can be used to demonstrate that the Unified Growth Theory is contradicted by data. The advantage of using the new compilation (Maddison, 2010) is that it helps to understand the recent transitions to slower trajectories because the earlier compilation was extended to include the data for the 21st century.

**Method of analysis and related issues**

We shall use two ways of displaying data: (1) semilogarithmic display of the GDP data and (2) the display of their reciprocal values, 1/GDP. These two types of display are suitable for studying data varying over a large range of values. The GDP values will be expressed in billions of 1990 International Geary-Khamis dollars.

Hyperbolic distributions, which describe the historical economic growth (Nielsen, 2016a), are represented by the simple mathematical formula:

$$S(t) = (a - kt)^{-1} \tag{1}$$

where, in our case, $S(t)$ is the GDP while $a$ and $k$ are positive constants.

The reciprocal values of hyperbolic distributions are represented by straight lines:

$$\frac{1}{S(t)} = a - kt \tag{2}$$

In general, hyperbolic growth can be uniquely identified by the decreasing straight line of the reciprocal values of the size of the growing entity in much the same way as the exponential growth can be identified by their logarithm. Reciprocal values of data can also help in identifying easily any deviations from hyperbolic trend because deviations from a straight line are easy to notice.

In using the reciprocal values it should be remembered that a deviation to a slower trajectory is indicated by an *upward* bending away from the previous linear trend while deviations to faster trajectories are indicated by *downward* bending. In particular, any form of boosting or takeoff, repeatedly claimed by Galor for global and regional economic growth, should be indicted by a clear change in the *downward* direction of the reciprocal values.

If the straight line fitting the reciprocal values of data remains undisturbed, it shows that there was no diversion to a faster or slower trajectory. In particular, if the straight line does not show a change in the downward direction (if the gradient of the trajectory of the reciprocal values remains constant) then there was no boosting in the economic growth. We obviously cannot claim a change of direction on an undisturbed straight line.

If the reciprocal values of data follow a decreasing straight line, the growth is not stagnant but hyperbolic. However, the concept of stagnation is not supported even if the reciprocal values of data do not decrease linearly. Any monotonically-decreasing trajectory will show that the postulate of stagnation followed by a takeoff at a certain time is not supported by data.

To prove the existence of the epoch of stagnation it is necessary to prove the presence of random fluctuations often described as Malthusian oscillations. Such random fluctuations should be clearly seen not only in the direct display of data but also in the display of their reciprocal values. It they are absent then there is no support in data for claiming the existence of the epoch of stagnation. Furthermore, if data do not show a clear takeoff from stagnation to growth at the postulated time, then there is no support for Galor's repeatedly-claimed takeoffs. However, if the reciprocal values of data follow a decreasing straight line, then they show, or at least strongly suggest, that the growth was hyperbolic.

If the straight line representing the reciprocal values of data remains unchanged, then obviously there is no change in the mechanism of growth. It is impossible to divide a straight line into different sections and claim different mechanism of growth for each of such arbitrarily selected sections. It is impossible to claim, for instance, a transition from stagnation to growth as repeatedly claimed by Galor in his Unified Growth Theory if the reciprocal values of data follow an undisturbed straight line. It is impossible to claim the existence differential takeoffs if there were no takeoffs. It is also impossible to claim that the Industrial Revolution changed the economic growth trajectory if the reciprocal values of data demonstrate that there was no change, i.e. that their linear trend remained undisturbed.

No-one has yet demonstrated the existence of Malthusian stagnation in the economic growth or in the growth of human population. For instance, Lee pointed out that "these models of Malthusian oscillations" are speculative when applied to the growth of human population (Lee, 1997, p. 1097). However, from the descriptions of Malthusian stagnation, its signature and the alleged escape from the Malthusian trap should be easy to identify. This signature is schematically presented in Figures 2 and 3.

For the direct display of GDP data (Figure 2), the signature of the regime of Malthusian stagnation can be identified by random fluctuations or oscillations around an approximately horizontal line. Over much longer sections of time, perhaps extending over thousands of years, fluctuations around the horizontal line might be replaced by fluctuations around a certain irregular trajectory (increasing, decreasing or randomly oscillating), which would be probably difficult to describe mathematically because the general concept of Malthusian stagnation is that it was controlled by random forces. Such random forces are hardly expected to generate monotonically-increasing distributions (Artzrouni and Komlos, 1985; Lagerlöf, 2006; McKeown, 2009; Komlos, 1989; van de Kaa, 2008). For the monotonically-increasing distributions, random forces are either too weak or they average out (Kapitza, 2006) and the growth is controlled by a certain dominant force, which could be constant (for the exponential growth), increasing with time or with the size of the growing entity (as for the hyperbolic growth) or even decreasing (as for the logistic growth).

The signature of the "remarkable" or "stunning" escape from the Malthusian trap (Galor, 2005a, pp. 177, 220) to the sustained economic growth should be easily identified by a clear takeoff from the earlier stagnant distribution to a fast increasing growth. The alleged escape should occur around AD 1750 for developed regions and around AD 1900 for less-developed regions (Galor, 2008a, 2012a).

For the reciprocal values of data (Figure 3), the epoch of Malthusian stagnation can be again identified by random fluctuations around an approximately horizontal line or around an irregular trajectory but the escape from the Malthusian trap will be identified by a clear *downward* trend. It should be noted that in the display of the reciprocal values of GDP data, small fluctuations are magnified, which means that in this display, epoch of Malthusian stagnation should be easy to identify because it should be characterised by strong fluctuations.

Maddison's data are indispensable in studying the historical economic growth but they have a strongly-limited range because they contain a large gap between AD 1 and 1000, and between AD 1000 and 1500. The most useful sets of data are from AD 1500. However, this shortcoming is immaterial because all the action described by Galor's three regimes of growth takes place after AD 1500. Within the range of the good sets of data, i.e. commencing from AD 1500, we should see clearly all the hallmarks of Galor's postulate of the three regimes of growth. We should see the signature of the regime of Malthusian stagnation, the effects of the Industrial Revolution, which was supposed to have been "the prime engine of economic growth" (Galor, 2005a, p. 212), the signature of the escape from the alleged Malthusian trap and a clear evidence of the uninterrupted era of the fast-increasing and sustained economic growth after stagnation. All these features should be clearly displayed. If they are not, then there is no support in the data for Galor's interpretations of the historical economic growth based on such distorted presentations of data as shown in Figure 1. Such presentations have no place in the scientific research.

The discussion presented here is the extension of the mathematical analysis of the historical economic growth (Nielsen, 2016a). We have already demonstrated that the historical economic growth was hyperbolic and thus that implicitly it gives no support for the doctrine of the three regimes of growth. Now, we shall show it explicitly.

It is essential to understand the fundamental features of hyperbolic distributions (Nielsen, 2014). Hyperbolic growth *is* slow over a long time and fast over a short time, but it is still the same, monotonically-increasing distribution, which is *impossible* to divide into two or three different, mathematically-justified components. The easiest way to see it is by using the reciprocal values [see the eqn (2)] because the confusing hyperbolic growth is then represented by a decreasing straight line. It is then clear that it is impossible to divide such a straight line into distinctly different, mathematically-justified components and claim distinctly different mechanisms of growth for each of these arbitrarily selected components.

Even though hyperbolic growth is slow over a long time it is *not* stagnant. *Slow hyperbolic growth should never be interpreted as stagnant* because if we want to interpret the slow perceived part of hyperbolic growth as stagnant, and governed by the usually assumed multitude of random forces, we should use precisely the same mechanism to explain the perceived fast component. The perceived slow and fast components belong to the same, monotonically-increasing distribution. It is impossible to divide a monotonically-increasing hyperbolic distribution into the mathematically-justifiable slow and fast sections because it is obviously impossible to divide a straight line describing the reciprocal values and representing the hyperbolic distribution into distinctly-different and mathematically-justifiable sections (Nielsen, 2014). It is scientifically unjustified to use different mechanisms of growth for such arbitrarily selected sections. Hyperbolic distributions have to be interpreted as a whole and the same mechanism has to be applied to the apparent slow growth and to the apparent fast growth. There is no clearly defined transition between the apparent slow and the apparent fast growth.

These comments apply also to the income per capita distributions represented by the Gross Domestic Product per capita (GDP/cap). Such distributions are even more confusing than hyperbolic distributions. They are linearly-modulated hyperbolic distributions, i.e. the monotonically-increasing hyperbolic distributions representing the growth of the GDP modulated by the monotonically-decreasing linear distributions representing the reciprocal values of the size of the population (Nielsen, 2015b). A product or a ratio of monotonic distributions cannot generate a non-monotonic distribution.

Even though the GDP/cap distributions appear to be made of two or maybe even three different components, as claimed incorrectly by Galor, they are increasing *monotonically* and it is impossible to divide them into distinctly different, mathematically-justifiable components. We can demonstrate it by calculating gradients or the growth rates of the GDP/cap distributions and by showing that they increase monotonically (Nielsen, 2015b). Any attempt to divide the GDP/cap distributions into distinctly-different components is strongly subjective and mathematically unjustified.

**Analysis of data for Western Europe**

We shall analyse two sets of data for Western Europe: (1) the data for 12 selected countries and the data for the total of 30 countries. The 12 selected countries are made of Austria, Belgium, Denmark, Finland, France, Germany, Italy, the Netherlands, Norway, Sweden, Switzerland and the United Kingdom. According to Maddison (2010), in 2008, these 12 countries accounted for 85% of the total GDP of the 30 countries of Western Europe. The total of the 30 countries includes also Ireland, Greece, Portugal, Spain and 14 other small west European countries.

The reason for analysing these two groups separately is that the listed 12 countries represent the most advanced economies, where the effects of the Industrial Revolution and the escape from the Malthusian trap should be most clearly visible. Consequently, for these 12 countries we should expect the best agreement between the Unified Growth Theory and the data.

Economic growth between AD 1 and 2008 in the 12 countries of Western Europe is shown in Figures 4 and 5. The growth in the total of 30 countries is shown in Figures 6 and 7.

Hyperbolic parameters describing economic growth in the 12 countries of Western Europe are: $a = 1.147 \times 10^{-1}$ and $k = 5.961 \times 10^{-5}$. The corresponding singularity is in 1923 but the economic growth was diverted to a slower trajectory around 1900, bypassing the singularity by about 23 years.

Hyperbolic fit to the data is remarkably good between AD 1500 and 1900 and acceptable below AD 1500. The point at AD 1 is only 27% higher than the fitted distribution and the point at AD 1000 is 54% lower. The critical range of time for testing the Unified Growth Theory is from AD 1500. It is in this range of time that we should be able to see transition from stagnation to growth and later a transition to the alleged sustained growth regime.

The data presented in Figures 4 and 5 clearly demonstrate that there is no support for the existence of the alleged regime of Malthusian stagnation. However, there is a convincing support for the hyperbolic growth at least between AD 1500 and 1900, the range of time where the signature of Malthusian stagnation should be still clearly displayed for about 300 years. The data show that during that time economic growth was following a steadily-increasing hyperbolic trajectory. There is no sign of the existence of Malthusian stagnation.

Absolutely nothing had happened at the end of the alleged Malthusian regime. There was no transition from stagnation to growth at any time. On the contrary, around the beginning of the postulated regime of sustained-growth, when the economic growth was supposed to have been launched from stagnation to a fast-increasing trajectory, the growth started to be diverted to a slower trajectory.

It is remarkable also that the Industrial Revolution had absolutely no impact on shaping the economic growth trajectory in these 12 countries. They should experience the greatest benefits of this revolution and they probably did but these benefits did not boost the economic growth. Technological innovations were used to sustain and propel economic growth but they did not change in the slightest the economic growth trajectory. In countries, where effects of the Industrial Revolution, "the prime engine of economic growth" (Galor, 2005a, p. 212), should have been most clearly reflected in the relevant data, we see no impacts of this engine.

This is an interesting issue, which should be studied and explained but it is futile to look for its explanation in the Unified Growth Theory. This interesting feature has not been even noticed by Galor, which is hardly surprising because it is hard or even impossible to carry out scientific research and draw reliable and scientifically-justified conclusions by repeatedly distorting data in such a way as shown in Figure 1.

Galor's Unified Growth Theory has no relevance to the description, let alone to the explanation of the mechanism of the economic growth, even in countries where his theory should be best fitted. Here, in the leading countries of Western Europe, where the effects of the Industrial Revolution should be most prominently displayed in the data describing economic growth, where the "remarkable" and "stunning" escape from the Malthusian trap (Galor, 2005a, pp. 177, 220) should be remarkably obvious, there are no signs of the impacts of the Industrial Revolution on the economic growth and no signs of any escape from the Malthusian trap, remarkable or less-remarkable, because there was no trap. Economic growth was increasing undisturbed and unconstrained along a hyperbolic trajectory until around 1900 when it started to be diverted to a slower but still fast-increasing trajectory.

Galor's three regimes of growth are totally dissociated from reality. They describe events that never happened.

Stories and explanations presented by Galor in his theory have no relevance to the explanation of the mechanism of the economic growth even in these 12 leading countries of Western Europe. His stories might be explaining or describing something else, e.g. social conditions or the style of living, but even then one wonders about the degree of reliability of such descriptions. His narrative does not explain the mechanism of the economic growth.

Results of the analysis of the economic growth in the total of 30 countries of Western Europe are presented in Figures 6 and 7. Hyperbolic parameters are: $a = 9.859 \times 10^{-2}$ and $k = 5.112 \times 10^{-5}$. The corresponding singularity is in 1929 but the economic growth was diverted to a slower trajectory around 1900, bypassing the singularity by about 29 years. The point at AD 1 is 42% higher than the calculated hyperbolic distribution and at AD 1000 it is 48% lover.

The analysis of the economic growth in the total of 30 countries of Western Europe leads to the same conclusions as for the 12 leading countries: Unified Growth Theory is contradicted by the economic growth data in Western Europe where the effects discussed by Galor should have been most convincingly confirmed. In contrast, they are convincingly contradicted.

**Analysis of data for Eastern Europe**

Results of the analysis of economic growth in Eastern Europe, based on using Maddison's data (Maddison, 2010), are presented in Figures 8 and 9. Hyperbolic parameters fitting the data are: $a = 7.749 \times 10^{-1}$ and $k = 4.048 \times 10^{-4}$. The point at AD 1 is 51% higher than the calculated curve. The singularity is in 1915 but the economic growth was diverted to a slower trajectory around 1890, bypassing the singularity by 25 years.

Unified Growth Theory is clearly contradicted by the economic growth data for Eastern Europe. The epoch of Malthusian stagnation did not exist within the range of the mathematically-analysable data. Outside of this range, any claim about the existence of the regime of Malthusian stagnation and about its effects on the economic growth has to be based on questionable conjectures. Such a claim would be also in conflict with the analysable data.

The data show no transition from stagnation to growth at any time because the growth was hyperbolic. There was no "remarkable" or "stunning" escape from the Malthusian trap (Galor, 2005a, pp. 177, 220) because there was no trap. Industrial Revolution did not boost the economic growth in Eastern Europe.

There was also no boosting of the economic growth at the time of the transition from the alleged post-Malthusian regime to the alleged sustained growth regime. Soon after the commencement of this phantom sustained-growth regime, economic growth in Eastern Europe started to be diverted to a slower trajectory. Galor's regimes of growth are clearly dissociated from data. They do not describe the real world but the world of fancy created by preconceived ideas and supported by the habitually-distorted presentation of data (Ashraf, 2009; Galor, 2005a, 2005b, 2007, 2008a, 2008b, 2008c, 2010, 2011, 2012a, 2012b, 2012c; Galor and Moav, 2002; Snowdon & Galor, 2008).

**Analysis of data for Asia**

Asia (excluding Japan) is made primarily, if not exclusively, of less-developed countries (BBC, 2014; Pereira, 2011). According to Galor, this region should have experienced the epoch of stagnation until around 1900 followed by the post-Malthusian regime commencing around that year. If Galor's claims are correct, we should see clear signs of stagnation in the data until around 1900 and a clear transition (a dramatic takeoff) from stagnation to growth around that year.

Economic growth in Asia between AD 1 and 2008 is presented in Figure 10. There is absolutely no correlation between the data and the three key events indicated in this figure: the Industrial Revolution, the alleged Malthusian regime and the alleged post-Malthusian regime, which were supposed to have been shaping the economic growth.

During the alleged Malthusian regime of stagnation, economic growth in Asia was increasing hyperbolically at least from AD 1000 but the point at AD 1 is also not far from the calculated hyperbolic distribution. Parameters fitting the data are $a = 2.303 \times 10^{-2}$ and $k = 1.129 \times 10^{-5}$.

The data show no signs of stagnation within their mathematically-analysable range, no signs of the Malthusian steady-state equilibrium and no signs of Malthusian oscillations. Assuming the existence of all such features is not only unnecessary but also scientifically unjustified because in science complicated interpretations are rejected in favour of simpler explanations. The data follow a steadily-increasing hyperbolic distribution, suggesting a simple mechanism of growth because hyperbolic distributions are described by a simple mathematical formula [see the eqn (1)].

The concept of stagnation is dramatically contradicted by data and so is the transition to the alleged post-Malthusian regime, which was supposed to have been a transition from stagnation to growth. We see no such transition but a continuation of the hyperbolic growth. The claimed by Galor takeoff did not happen. There was a minor and hard-to-notice disturbance in the economic growth around 1950 but the growth soon returned to its historical hyperbolic trajectory. The overall evidence in the data is that the propping-up structures (the alleged different regimes of growth) used by Galor are not only totally redundant but also strongly misleading. They can, and even should, be removed because the data reveal a totally different pattern of growth.

The data and their analysis show that nothing dramatic occurred during the alleged transition from the postulated Malthusian regime of stagnation to the alleged post-Malthusian regime, which is supposed to mark the escape from the postulated Malthusian trap and leading to a sustained growth regime. There was no escape from the trap because there was no trap. During the postulated Malthusian trap the economic growth was steadily increasing and it was obviously unconstrained. It is futile to claim random fluctuations and oscillations when there are none. Why should we even contemplate to make it all more complicated when the data show that the growth was much simpler?

If not for Maddison and his data, the established knowledge in the economic research would have remained established, but now it has to be revaluated and changed. However, new insights should be welcome, particularly if they suggest a simpler explanation of the historical economic growth.

Reciprocal values of the GDP data, 1/GDP, shown in Figure 11, also demonstrate that the Unified Growth Theory is contradicted by the same data, which were used during its development, the data published by Maddison in 2001 (Maddison, 2001) but later extended to include economic growth during the 21st century (Maddison, 2010).

During the alleged Malthusian regime of stagnation, reciprocal values of data were decreasing along a straight line indicating an undisturbed, hyperbolic economic growth. The data show also that nothing dramatic had happened at the end of this alleged epoch of stagnation. There was no transition to a new regime of growth. In particular, there was no transition from stagnation to growth, as claimed by Galor, but a continuation of the hyperbolic growth. The concept of the two regimes of growth is convincingly contradicted by data.

**Analysis of data for the former USSR**

Economic growth in the countries of the former USSR between AD 1 and 2008 is presented in Figure 12. Reciprocal values of the GDP data, 1/GDP, are shown in Figures 13 and 14. The growth was hyperbolic between AD 1 and around 1870. Parameters describing hyperbolic growth are $a = 6.547 \times 10^{-1}$ and $k = 3.452 \times 10^{-4}$.

During the entire range of the mathematically-analysable data the epoch of Malthusian stagnation did not exist. Galor's regimes of growth are hanging there without having any connection with data. The "remarkable" or "stunning" escape from the Malthusian trap did not happen because there was no trap. Galor's Malthusian regime ends in the middle of nowhere. Absolutely nothing (remarkable or less-remarkable, stunning or less stunning) happened on the border between the alleged Malthusian regime and the post-Malthusian regime. There was also no stunning or remarkable escape at the onset of the alleged sustained-growth regime. There was no dramatic increase in the economic growth. On the contrary, economic growth started to be diverted to a slower trajectory.

What is remarkable about the confrontation of Galor's theory with the empirical evidence is that there is such a consistently repeated and stunning disagreement between his theory and the data. The data also demonstrate that the Industrial Revolution had absolutely no impact on changing the economic growth trajectory in the countries of the former USSR. Here again we see that "the prime engine of economic growth" (Galor, 2005a, p. 212) did nothing to change to growth trajectory. Whatever this engine might have been doing, it certainly did not boost the economic growth. The data and their analysis give no support for the concept of Malthusian stagnation and for the assumption of the existence of the steady-state Malthusian equilibrium. Economic growth was increasing along a remarkably-stable hyperbolic trajectory. There was no escape from the Malthusian trap, let alone a "remarkable" or "stunning" escape as claimed by Galor (2005a, pp. 177, 220), because there was no trap. The growth was always unconstrained because the hyperbolic trajectory remained unimpeded.

The concept of stagnation is dramatically contradicted by data and so is the alleged transition from stagnation to growth. Such a transition never happened. On the contrary, from around 1870, economic growth in the countries of former USSR started to be diverted to a slower trajectory, away from its faster, historical hyperbolic trajectory.

**Analysis of data for Africa**

Africa is a perfect example of a cluster of countries, which belong to the group of less-developed and least-developed countries. Out of the total of 48 least-developed countries in the world, 34 are in Africa (Bangla News, 2015; UNCTAD, 2013). With just one minor exception, Africa is made entirely of less-developed and least-developed countries (BBC, 2014; Pereira, 2011). The exception is Western Sahara, a small country in transition made of around 586,000 people (UNDATA, 2015).

Maddison's data for Africa serve, therefore, as an excellent source of information to test Galor's hypothesis of the existence of the distinctly different regimes of economic growth in less-developed regions. We shall demonstrate that this hypothesis is dramatically and clearly contradicted by data.

Reciprocal values of data describing economic-growth in Africa are presented in Figure 15. Economic growth was clearly hyperbolic between AD 1 and around 1820 because the reciprocal values follow a straight line. There was definitely no stagnation. The concept of the regime of Malthusian stagnation is clearly contradicted by data. To prove its existence one would have to demonstrate a stagnant state of growth characterised by random Malthusian oscillation around an approximately horizontal line as shown in Figure 3. The data contain no such signature. On the contrary they show a steadily-increasing and remarkably-stable hyperbolic growth. There are no signs of any possible fluctuations, which in this representation of data should be strongly magnified.

Furthermore, Galor's concept of Malthusian stagnation extending to 1900 ignores not only the data between AD 1 and 1820 but also the clear and dramatic transition, which occurred around 1820. It was *not* a transition from stagnation to growth but *from growth to growth*, the transition from a slower but steadily-increasing hyperbolic growth to a faster and steadily-increasing hyperbolic growth. This pattern is in clear contradiction of the Unified Growth Theory (Galor, 2005a, 2008a, 2011, 2012a).

The concept of the regime of stagnation ignores the steadily-increasing economic growth before 1820, the dramatic change in the pattern of growth around that year and the new hyperbolic growth after 1820. The claim of Malthusian stagnation ending in 1900 for less-developed countries ignores also that absolutely nothing unusual had happened around that year. The economic growth continued undisturbed. The postulated Malthusian regime ends in the middle of nowhere. There is no justification for claiming the regime of Malthusian stagnation and no justification for terminating it in AD 1900 or at any other time because there was no stagnation.

In addition, the data demonstrate the existence of a feature, which is ignored by Galor: the diversion to a slower trajectory around 1950 indicated by the upward bending of the trajectory of the reciprocal values. According to Galor, the economic growth was supposed to have been boosted from stagnation to growth (at the end of his alleged Malthusian regime) and launched into a fast-increasing growth, but data present an entirely different interpretation: economic growth was increasing fast along a hyperbolic trajectory during the alleged regime of Malthusian stagnation but shortly after the time of the postulated transition to a faster growth the data started to follow a *slower* trajectory. Data tell one story, Galor tells another, and in science data have the priority.

The disagreement between Galor's theory and the data is also clearly demonstrated in Figures 17 and 18. Over the range of the mathematically-analysable data the Malthusian regime did not exist. The data show no evidence of the features characterising the epoch of Malthusian stagnation. In contrast, the data show steadily-increasing hyperbolic distributions.

In his description of economic growth, Galor did not even notice that there was a strong transition around AD 1820, let alone that it was a transition from one hyperbolic distribution to another. He also did not notice that that the postulated epoch of Malthusian stagnation ends in the middle of nowhere (see Figure 18).

Many important details are easily lost in the habitually distorted presentations of data (Ashraf, 2009; Galor, 2005a, 2005b, 2007, 2008a, 2008b, 2008c, 2010, 2011, 2012a, 2012b, 2012c; Galor and Moav, 2002; Snowdon & Galor, 2008) as illustrated in Figure 1. It is hard or even impossible to draw reliable conclusions by using such distorted diagrams and by making no attempt to analyse data. Conclusions based on impressions are likely to be incorrect. It is hard or even impossible to do science without following the principles of scientific investigation.

**Analysis of data for Latin America**

Results of analysis of the economic growth in Latin America based on Maddison's data (Maddison, 2010) are shown in Figures 19 and 20.

The data suggest the existence of two hyperbolic growth trajectories: a slow trajectory between AD 1 and 1500 and a fast trajectory between AD 1600 and 1870. The slow trajectory is characterised by parameters $a = 4.421 \times 10^{-1}$ and $k = 2.093 \times 10^{-4}$. The singularity for this trajectory was at $t_s = 2113$. The fast trajectory is characterised by parameters $a = 1.570 \times 10^0$ and $k = 8.224 \times 10^{-4}$. The singularity for this new trajectory was at $t_s = 1910$. However, from around 1870, i.e. from around the time of the alleged takeoff from stagnation to growth (Galor, 2008a, 2012a), economic growth in Latin America started to be diverted to a slower trajectory bypassing the singularity by a safe margin of 40 years. The illusion of a *takeoff* is replaced by a diversion to a *slower* growth.

The characteristic features of the economic growth in Latin America are similar to the features in Africa. In both cases, a slow hyperbolic growth was followed by a much faster hyperbolic trajectory and this transition can be correlated with the intensified colonisation of Latin America (Bethell, 1984).

The data for Latin America are in clear disagreement with the Unified Growth Theory. The economic growth was slow before AD 1500 but there is no basis for claiming that it was stagnant. Hyperbolic trajectory between AD 1 and 1500 could be questioned but it is consistent with the similar, but much clearer, pattern in Africa and is in perfect agreement with the repeated evidence of hyperbolic growth in other regions. There is definitely no convincing support for the existence of the epoch of stagnation.

The data show a brief economic decline between AD 1500 and 1600, which appears to be coinciding with the commencement of the intensified Spanish conquest (Bethell, 1984). However, from around AD 1600, economic growth in Latin America was following a fast-increasing hyperbolic trajectory. The change from a slow to fast economic growth occurred about *300 years before the alleged takeoff around 1900*. Furthermore, as in Africa, *it was not a transition from stagnation to growth but from hyperbolic growth to hyperbolic growth*. This feature is ignored in the Unified Growth Theory. Remarkably also, at the time of the alleged "remarkable" escape from the Malthusian trap (Galor, 2005a, p. 177) in around AD 1900, economic growth in Latin America was already diverted to a *slower* trajectory.

Unified Growth Theory presents a story, which is contradicted by data. There is no correlation between the data and the narrative of this theory. In his habitually crude display of data, Galor could not have seen all these important features. He appears to have been guided by the inherited ideas, which unfortunately he did not check by the rigorous analysis of the new and excellent data (Maddison, 2001) available to him at the time of the formulation of his theory. The updated compilation of the data describing the historical economic growth (Maddison, 2010) was also available to him even before the publication of his book (Galor, 2011) and certainly during his continuing dissemination of the same ideas after its publication. As mentioned earlier, any of these compilations can be used to show that Galor's theory is fundamentally incorrect because during the time when there were supposed to have been transitions between alleged regimes of growth the two compilation contain the same data and they show a clear disagreement with Galor's theory.

**Summary and conclusions**

We have analysed economic growth in Western Europe, Eastern Europe, Asia, former USSR, Africa and Latin America (Maddison, 2010). We have found that the fundamental concepts of the Unified Growth Theory (Galor, 2005a, 2011) are contradicted by the same data, which were used but never analysed during the formulation of this theory.

Whatever was wished-for did not happen. The real world refused to comply with the preconceived ideas and with the imagined interpretations, which were creating such an attractive story.

It seems to be obvious that the Industrial Revolution should have a strong and decisive effect on the economic growth but it did not. It seems to be obvious that a slow growth is stagnant but it is not. What seems to be obvious is not necessarily true. It is obvious that the Sun moves around the Earth but it does not. "It is clear that the earth does not move, and that it does not lie elsewhere than at the centre" (Aristotle).

Empirical evidence has to be methodically and carefully analysed; otherwise we shall be creating our own stories, which might be interesting, exciting and convincing but they will be stories of fiction. They will have nothing to do with science. In science we learn from nature. Any attempt to mould nature into the image fashioned by our creative imagination is bound to fail and the perfect example is the Unified Growth Theory.

Within the range of the mathematically-analysable data, the three regimes of growth, the Malthusian regime, the post-Malthusian regime and the sustained-growth regime did not exist. There is no correlation between the data and these three postulated regimes of growth. In particular, there was no escape from the Malthusian trap because there was no trap.

During the time described by the mathematically-analysable data, economic growth was hyperbolic and generally undisturbed. Only most recently, around the time when according to the Unified Growth Theory it should have been boosted from stagnation to growth, economic growth started to be diverted from the fast-increasing hyperbolic trajectories to slower trajectories. Unified Growth Theory does not explain, let alone describe the historical economic growth because it is based on the fundamentally incorrect premises.

The concept of the three regimes of growth was supported by the distorted presentation of data (Ashraf, 2009; Galor, 2005a, 2005b, 2007, 2008a, 2008b, 2008c, 2010, 2011, 2012a, 2012b, 2012c; Galor and Moav, 2002; Snowdon & Galor, 2008). When properly displayed and analysed, the same data show that the Unified Growth Theory is fundamentally incorrect.

The reliable and correct interpretation of the historical economic growth might appear to have no practical application because what was in the past is in the past. Why should the distant past have any influence on our present economic growth? However, the correct understanding of the past economic growth may well decide about our future.

Galor's interpretations of the historical economic growth are not only scientifically unacceptable but also dangerously incorrect because they create the false sense of security. They present a picture of the unsustained economic growth in the past and of a transition to a new era of sustained economy after the "remarkable" or "stunning" escape from the Malthusian trap (Galor, 2005, pp. 177, 220). At last, after the endless suffering, straggle, and deprivation, humans escaped the tyranny of the Malthusian regime and now they can enjoy the sustained economic growth with its prosperous future. This is a pleasing story but the opposite is true.

Rigorous analysis of data shows convincingly that the past economic growth was sustained and secure because it followed the remarkably stable hyperbolic trajectories (Nielsen, 2016a). This conclusion is in harmony with the study of ecological footprints, which shows that until the late 1900s global ecological footprint was lower than the ecological capacity (WWF, 2010). It was in the past that the economic growth was not only sustained but also sustainable. Now it is not, because it is supported by the increasing ecological deficit. Indeed, mathematical analysis of the economic growth shows that its future is insecure (Nielsen, 2015a).

Economic growth was not in a trap in the past but now it is in a trap of our continuing drive to increase not only the GDP but also the GDP/cap. We seem to see no limit to prosperity but the limit is imposed by the ecological limits and by the fast-increasing trajectories of economic growth. While the Unified Growth Theory suggests a prosperous future of the "sustained growth regime" after the alleged "Malthusian regime," the data indicate that unless we take decisive steps to control the current economic growth our future is insecure (Nielsen, 2015a).

In its present form, Galor's Unified Growth Theory is unacceptable. It has to be either thoroughly revised or rejected and replaced by a new theory aimed at explaining why the economic growth was hyperbolic in the past, why it was increasing along such remarkably stable trajectories, why it started to be diverted to slower, but still fast-increasing, trajectories and, most importantly, how to create a sustainable economic future.

Propelled by the gained momentum of the historical economic growth, the current growth continues to increase too fast. It has to be slowed down. The sustainable and secure economic growth has yet to be created. It has not been created automatically at the end of the alleged but non-existent Malthusian regime as suggested incorrectly by the Unified Growth Theory.

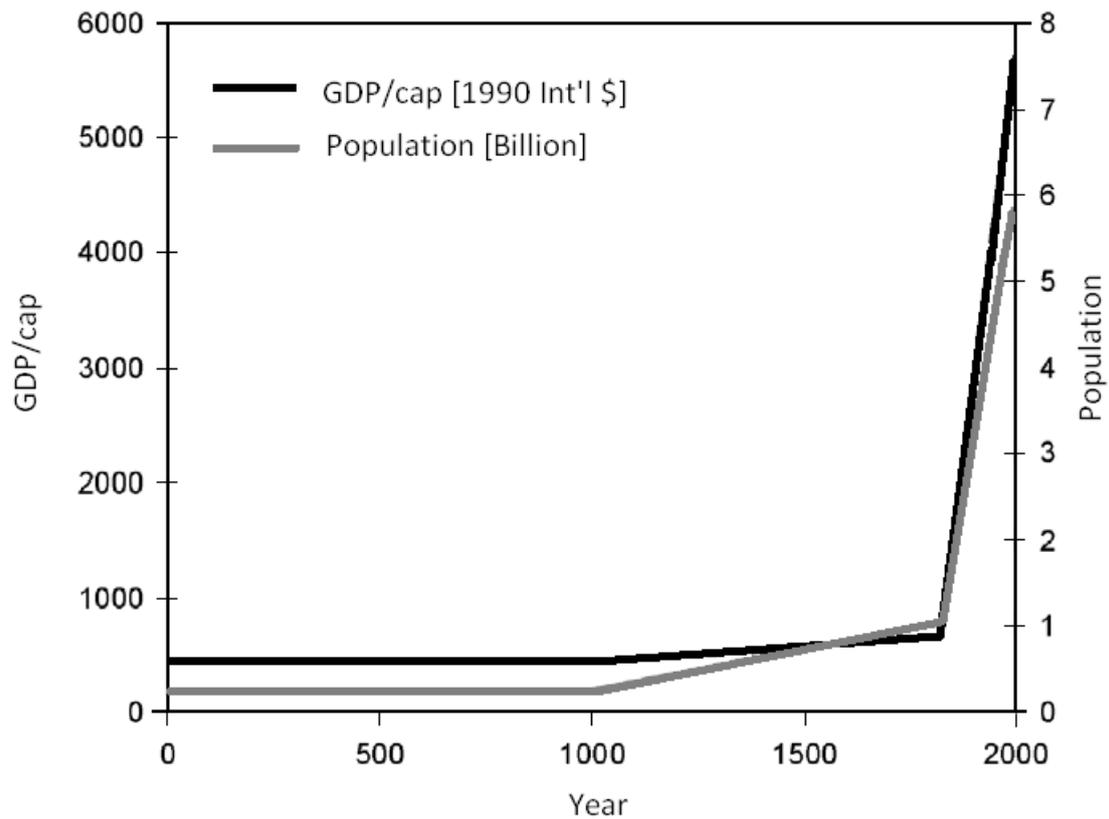

**Figure 1.** Example of the ubiquitous, grossly-distorted and self-misleading diagrams used to create the Unified Growth Theory (Galor, 2005a, 2011). Madison's data (Maddison, 2001) were used during the formulation of this theory but they were never analysed. Such a state-of-the-art of the presentation of data was used to construct a system of scientifically-unsupported concepts, interpretations and explanations.

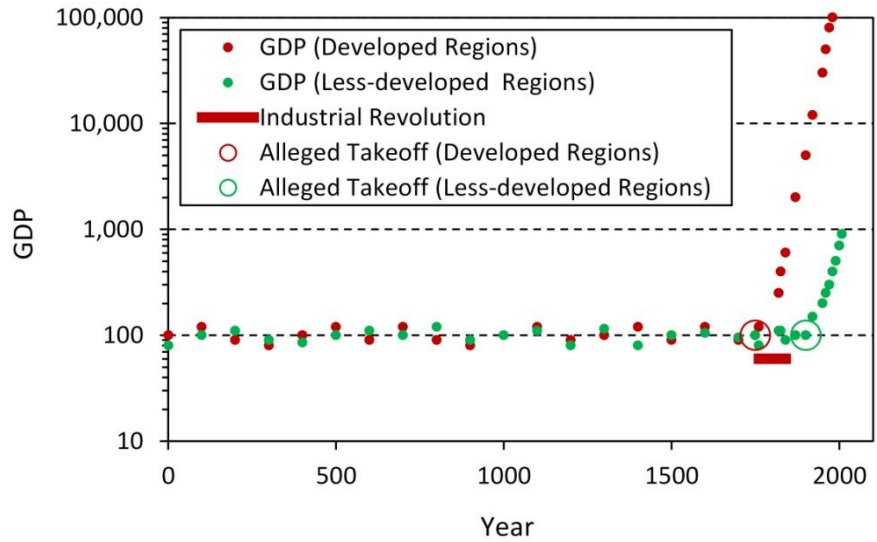

**Figure 2.** Direct display of the hypothetical GDP data serving as the schematic representation of the signature of Malthusian stagnation (fluctuations around an approximately horizontal line) followed by the escape from the Malthusian trap into the sustained economic-growth regime around AD 1750 for developed regions and around AD 1900 for less-developed regions as claimed by Galor (2005a, 2008a, 2011, 2012a). If these signatures are missing, Unified Growth Theory is contradicted by data.

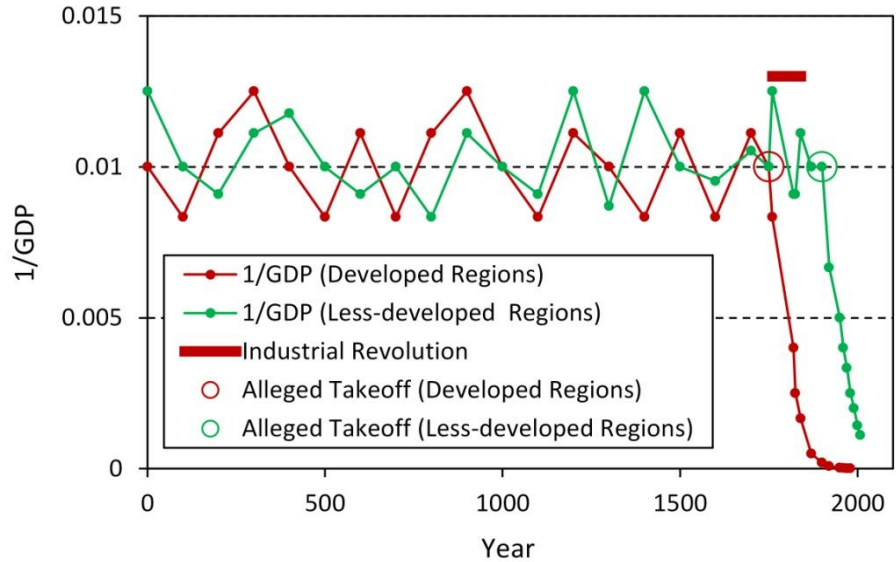

**Figure 3.** Display of the reciprocal values of the same hypothetical data as shown in Figure 2, serving as the schematic representation of the signature of Malthusian stagnation (fluctuations around an approximately horizontal line) followed by the escape from the Malthusian trap into the sustained economic-growth regime around AD 1750 for developed regions and around AD 1900 for less-developed regions as claimed by Galor (2005a, 2008a, 2011, 2012a). If these signatures are absent, Unified Growth Theory is contradicted by data.

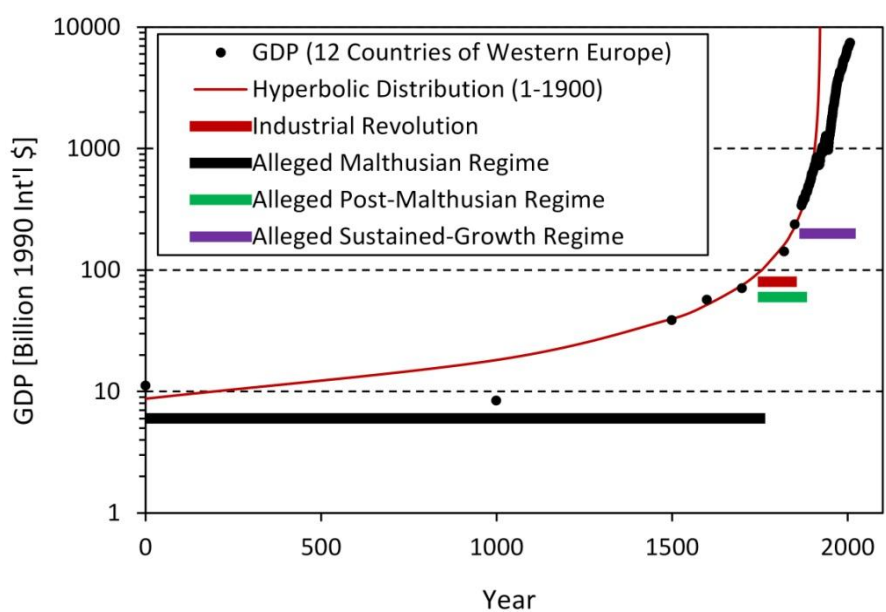

**Figure 4.** Economic growth in the 12 selected countries of Western Europe representing the most-advanced economies where the Unified Growth Theory should have the strongest confirmation. There was no transition from stagnation to growth at any time. The growth was hyperbolic before and after the alleged transition around AD 1750. Industrial Revolution did not boost the economic growth. The "remarkable" or "stunning" escape from the Malthusian trap (Galor, 2005a, pp. 177, 220) did not happen because there was no trap. Galor's three regimes of growth have no relevance to the description, let alone to the explanation, of the mechanism of the economic growth. During the alleged sustained growth regime, when the economic growth was supposed to follow a fast-increasing trajectory after the epoch of stagnation, economic growth was diverted to a *slower* trajectory.

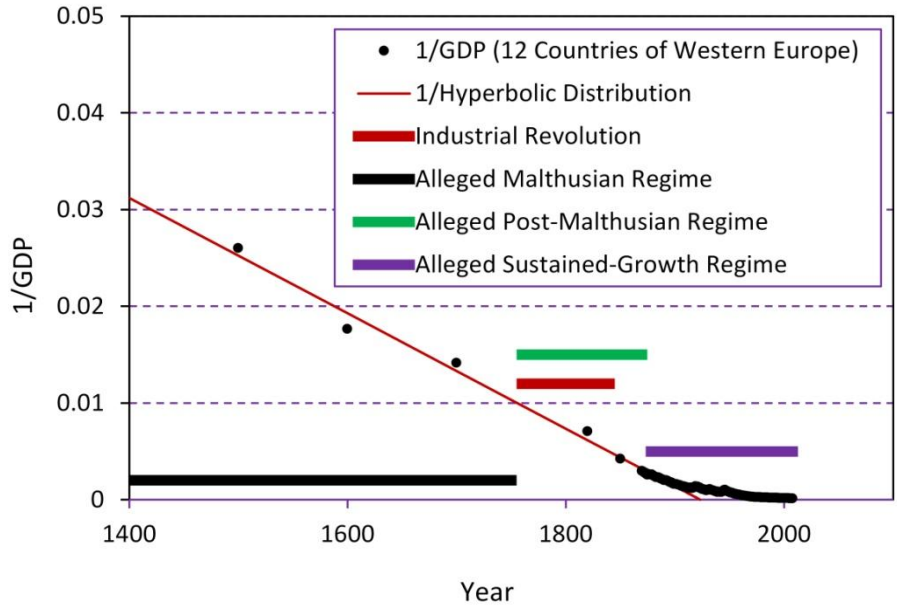

**Figure 5.** Reciprocal values of the GDP data, 1/GDP, for the economic growth in the 12 selected countries of Western Europe. Unified Growth Theory (Galor, 2005a, 2011) is contradicted by Maddison's data (Maddison, 2010). Galor's three regimes of growth have no relevance to the description, let alone to the explanation, of the mechanism of the economic growth. There was no transition from stagnation to growth at any time because there was no stagnation. There was no "remarkable" or "stunning" escape from the Malthusian trap (Galor, 2005a, pp. 177, 220) because there was no trap. Industrial Revolution did not boost the economic growth even in the countries where its effects should be most pronounced. During the alleged sustained growth regime, when the economic growth was supposed to follow a fast-increasing trajectory after the epoch of stagnation, economic growth was diverted to a *slower* trajectory, as indicated by the upward bending of the trajectory of the reciprocal values.

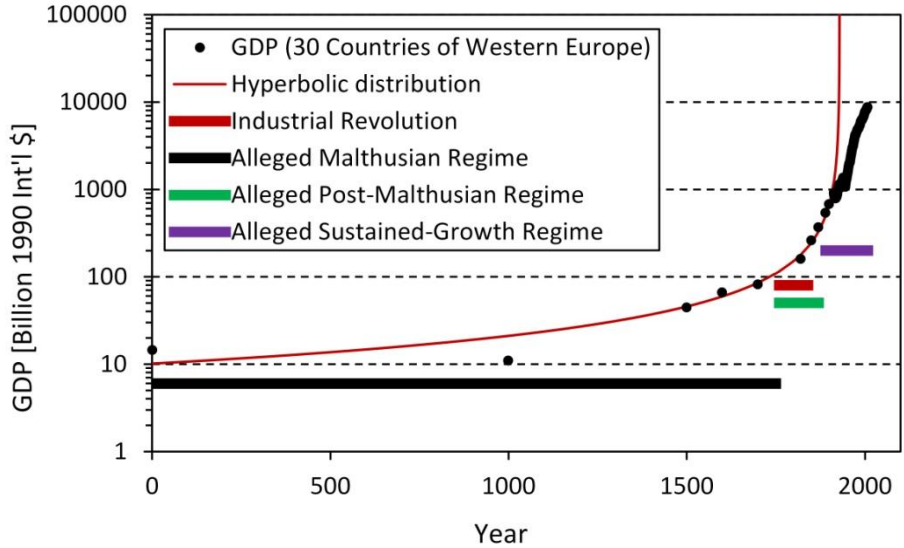

**Figure 6.** Economic growth in the total of 30 countries of Western Europe. The data give no clear support for the existence of the alleged Malthusian regime of stagnation. Industrial Revolution did not boost the economic growth in Western Europe. The "remarkable" or "stunning" escape from the Malthusian trap (Galor, 2005a, pp. 177, 220) did not happen because there was no trap. Galor's three regimes of growth have no relevance to the description or to the explanation of the mechanism of the economic growth in Western Europe. During the alleged sustained growth regime, when the economic growth was supposed to follow a fast-increasing trajectory after the epoch of stagnation, economic growth was diverted to a *slower* trajectory.

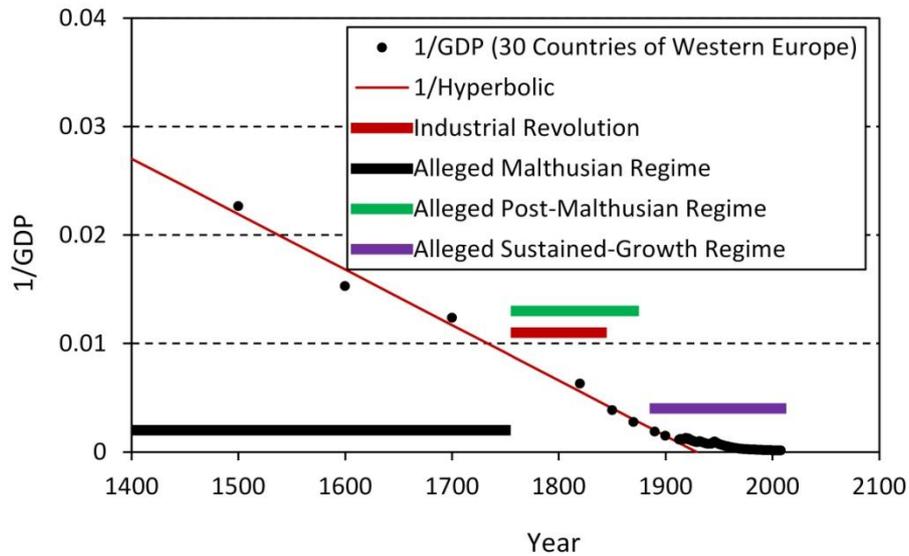

**Figure 7.** Reciprocal values of the GDP data, 1/GDP, for the economic growth in the total of 30 countries of Western Europe. Unified Growth Theory (Galor, 2005a, 2011) is contradicted by Maddison's data (Maddison, 2010). Galor's three regimes of growth have no expected correlation with data. There was no transition from stagnation to growth at any time because there was no stagnation. There was no "remarkable" or "stunning" escape from the Malthusian trap (Galor, 2005a, pp. 177, 220) because there was no trap. Industrial Revolution did not boost the economic growth in Western Europe. During the alleged sustained growth regime, when the economic growth was supposed to follow a fast-increasing trajectory after the epoch of stagnation, economic growth was diverted to a *slower* trajectory

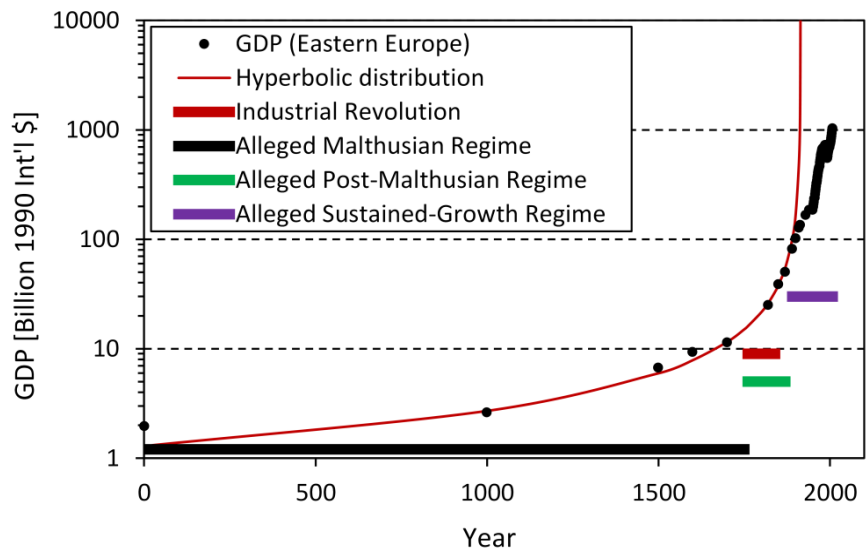

**Figure 8.** Economic growth in Eastern Europe. Galor's three regimes of growth have no relevance to the description, let alone to the explanation, of the mechanism of the economic growth. Unified Growth Theory is contradicted by data. The alleged Malthusian regime of stagnation did not exist. Industrial Revolution did not boost the economic growth in Eastern Europe. The "remarkable" or "stunning" escape from the Malthusian trap (Galor, 2005a, pp. 177, 220) did not happen because there was no trap. During the alleged sustained growth regime, when the economic growth was supposed to follow a fast-increasing trajectory after the epoch of stagnation, economic growth was diverted to a *slower* trajectory

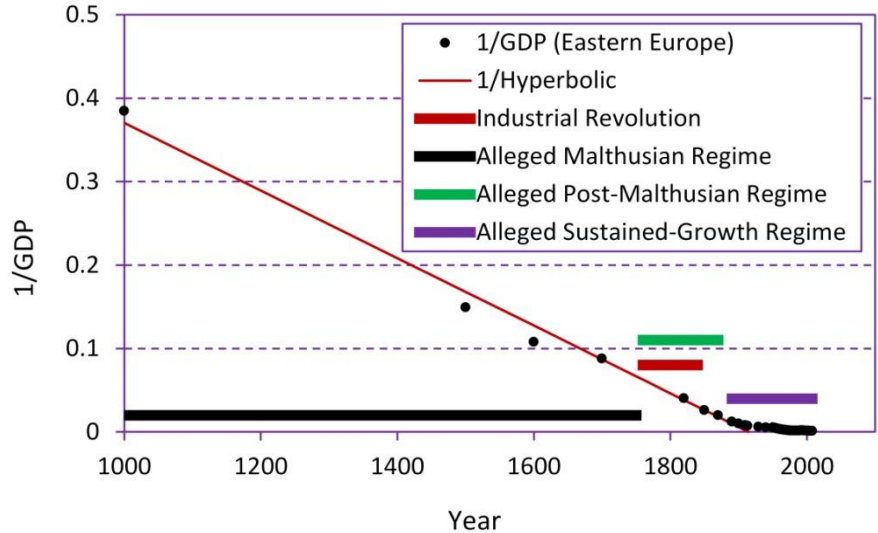

**Figure 9.** The reciprocal values of the GDP data, 1/GDP, for the economic growth in Eestern Europe. Unified Growth Theory (Galor, 2005a, 2011) is contradicted by Maddison's data (Maddison, 2010). Galor's three regimes of growth have no expected connection with data. There was no transition from stagnation to growth at any time because there was no stagnation. There was no "remarkable" or "stunning" escape from the Malthusian trap (Galor, 2005a, pp. 177, 220) because there was no trap. Industrial Revolution did not boost the economic growth in Eastern Europe. Galor's theory has no relevance to the description, let alone to the explanation, of the mechanism of the economic growth. During the alleged sustained growth regime, when the economic growth was supposed to follow a fast-increasing trajectory after the epoch of stagnation, economic growth was diverted to a *slower* trajectory, as indicated by the upward bending of the trajectory of the reciprocal values.

**Asia (excluding Japan)**

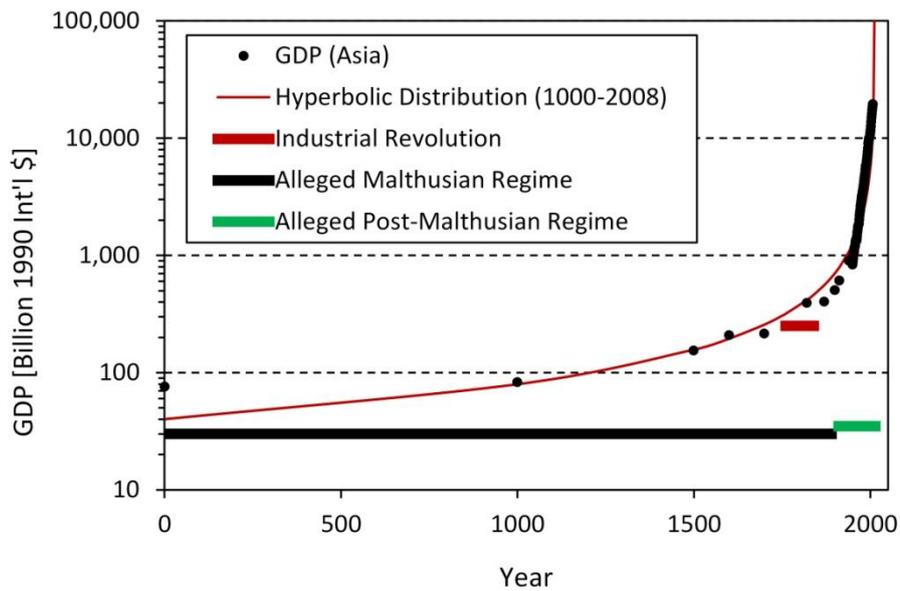

**Figure 10.** Economic growth in Asia (excluding Japan) between AD 1 and 2008. Maddison's data (Maddison, 2010) are compared with the hyperbolic distribution and with their unsubstantiated interpretations promoted by Galor (Galor, 2005a, 2011). Economic growth was hyperbolic from at least AD 1000 until 2008. The minor delay after the Industrial Revolution was followed by the compensating recovery. The concept of the alleged Malthusian regime of stagnation is contradicted by data. The escape from the Malthusian trap never happened because there was no trap. There was no dramatic transition from stagnation to growth because there was no stagnation.

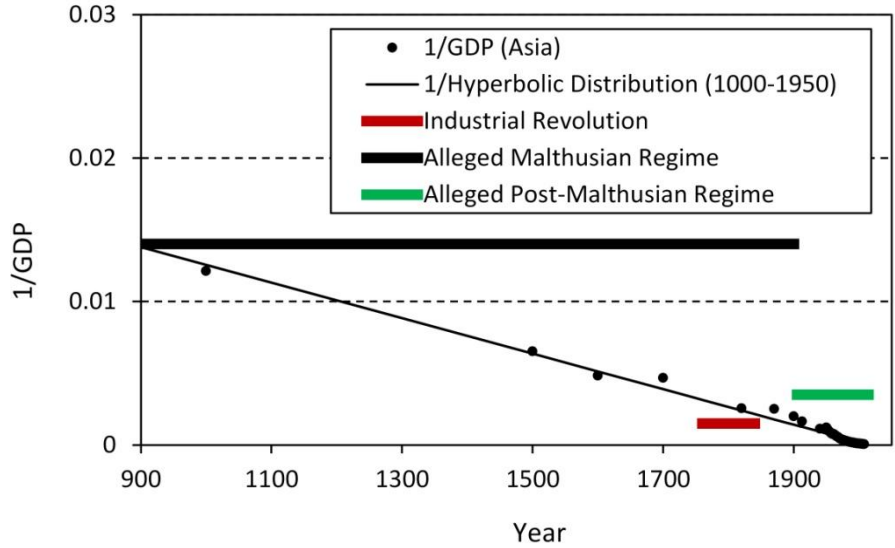

**Figure 11.** Reciprocal values of the GDP data, 1/GDP, for Asia demonstrate that there is no correlation between the claimed events (Industrial Revolution, the alleged Malthusian regime of stagnation and the alleged post-Malthusian regime) and the data (Maddison, 2010). The postulated dramatic and remarkable takeoff around 1900 never happened. The Malthusian regime of stagnation and the post-Malthusian regime did not exist.

**Former USSR**

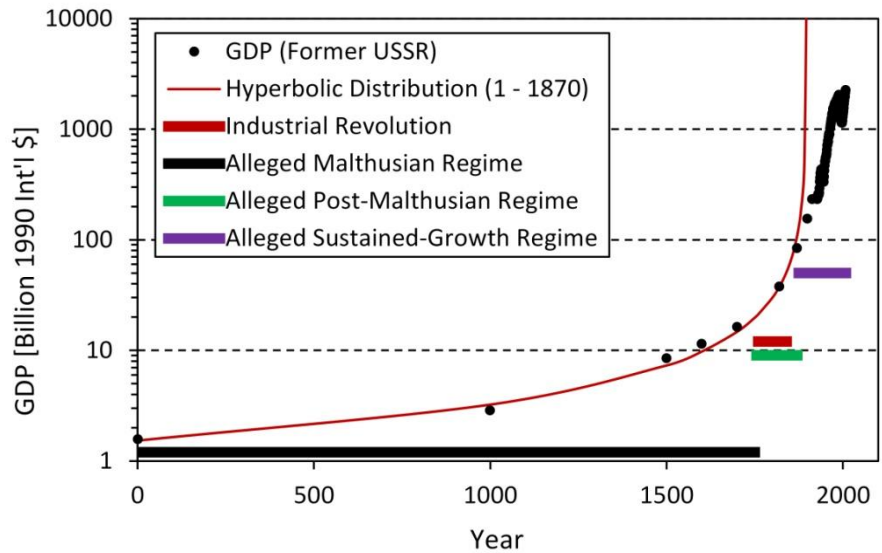

**Figure 12**. Economic growth in the countries of the former USSR between AD 1 and 2008, as represented by Maddison's data (Maddison, 2010), is compared with the hyperbolic distribution and with the unsubstantiated interpretations of the mechanism of growth proposed by Galor (Galor, 2005a, 2008a, 2011, 2012a). The alleged Malthusian regime of stagnation did not exist and neither did the alleged post-Malthusian and sustained-growth regimes. The Industrial Revolution had absolutely no impact on changing the economic growth trajectory. There was also no dramatic transition to a new and faster economic growth after the alleged epoch of stagnation, no transition from stagnation to growth at any time because there was no stagnation. There was no escape from the Malthusian trap because there was no trap. In place of all these imaginary and wished-for features there was the undisturbed and well-sustained hyperbolic growth. During the alleged sustained growth regime, when the economic growth was supposed to follow a fast-increasing trajectory after the epoch of stagnation, economic growth was diverted to a *slower* trajectory.

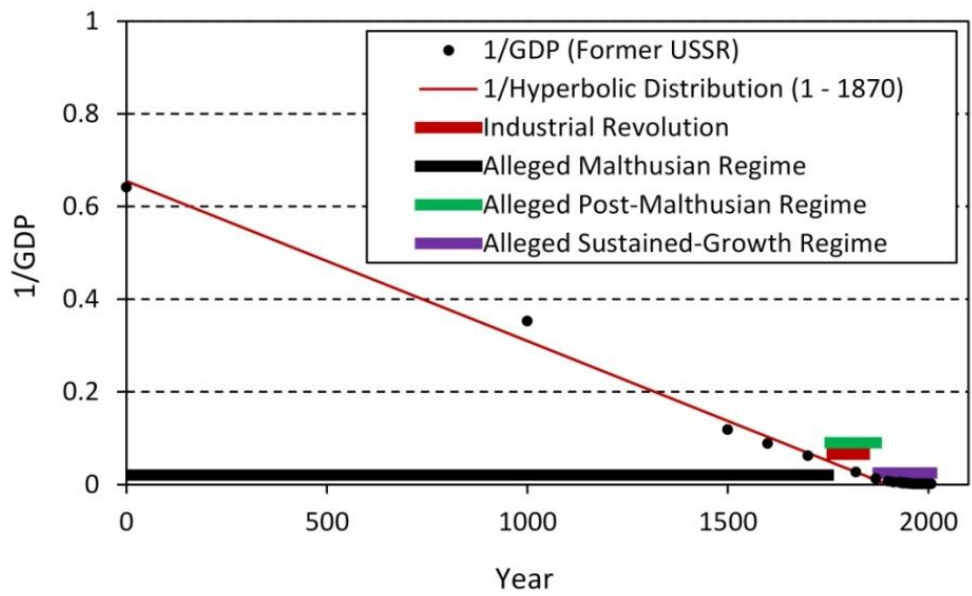

**Figure 13.** Reciprocal values of the GDP data, 1/GDP, for the former USSR are compared with the hyperbolic distribution represented by the decreasing straight line. There was no stagnation. Throughout the entire range of the alleged Malthusian regime during the AD era, economic growth was hyperbolic.

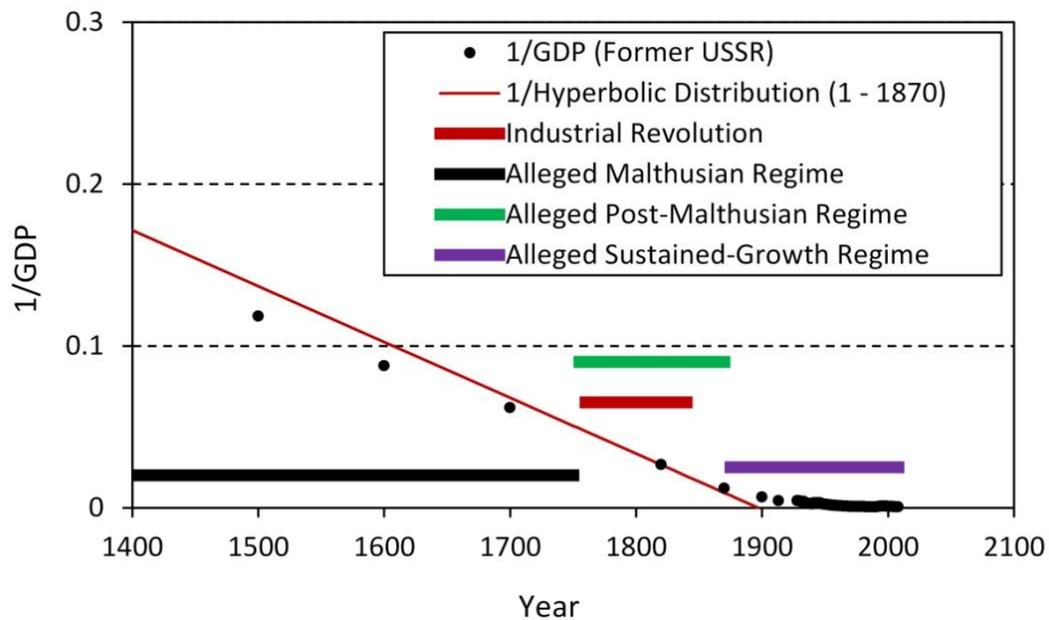

**Figure 14.** The end part of the plot of the reciprocal values of the GDP data, 1/GDP, for the former USSR. Economic growth was hyperbolic until around AD 1870 when it started to be diverted to a slower trajectory indicated by an upward bending of the reciprocal values. Industrial Revolution did not boost the economic growth. The alleged Malthusian regime of stagnation did not exist and there was no transition from stagnation to growth at any time because there was no stagnation. The "stunning" or "remarkable" escape from the Malthusian trap (Galor, 2005a, pp. 177, 220) did not happen because there was no trap. During the alleged sustained growth regime, when the economic growth was supposed to follow a fast-increasing trajectory after the epoch of stagnation, economic growth was diverted to a *slower* trajectory.

**Africa**

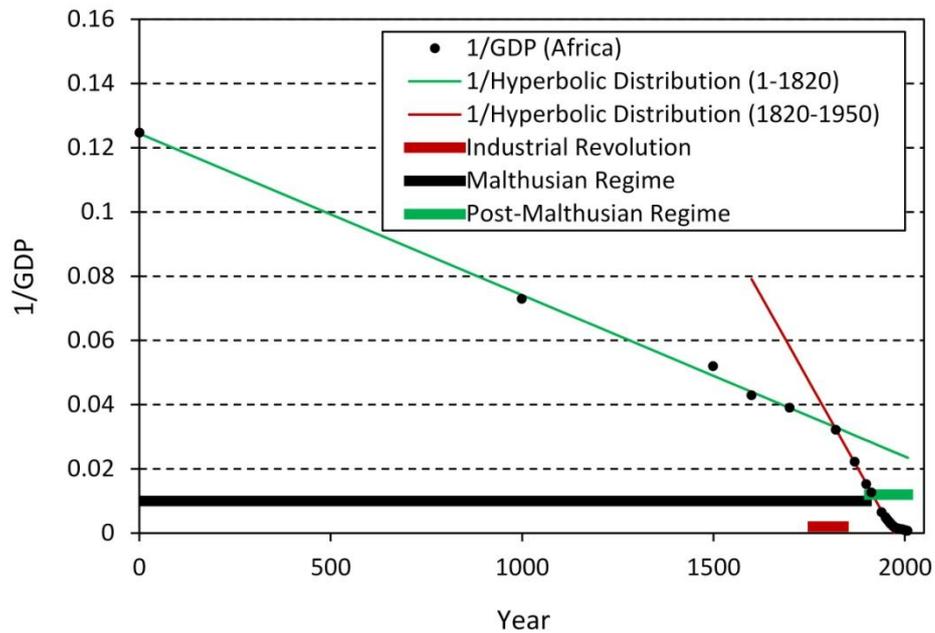

**Figure 15.** Reciprocal values of the GDP data (Maddison, 2010) for Africa compared with the hyperbolic distributions represented by the decreasing straight lines. The two distinctly different regimes of growth postulated by Galor (2005a, 2008a, 2011, 2012a) did not exist. There was no transition from stagnation to growth at any time because there was no stagnation.

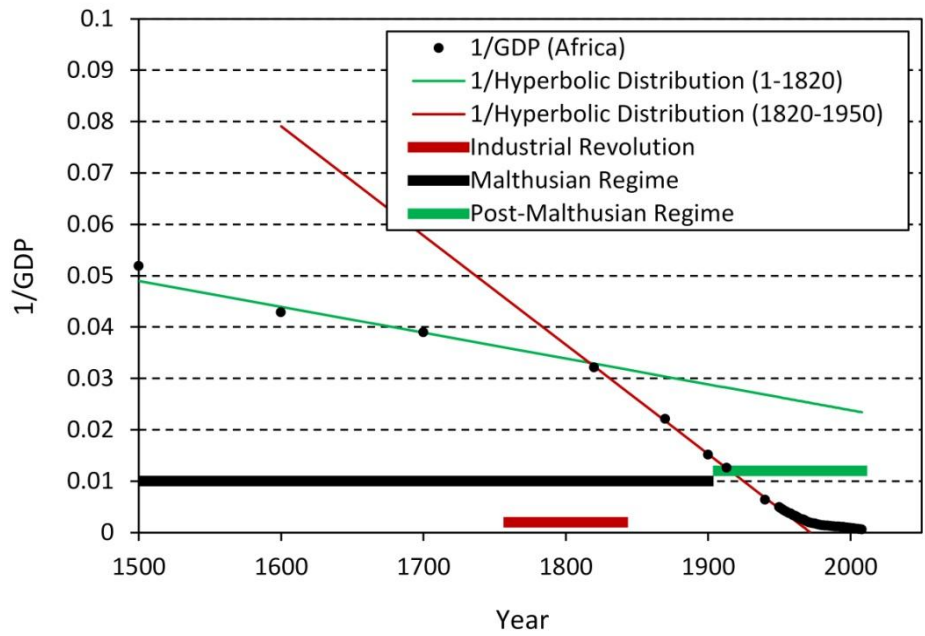

**Figure 16.** Reciprocal values of the GDP data (Maddison, 2010) for Africa between AD 1500 and 2008 compared with the hyperbolic distributions represented by the decreasing straight lines. The two distinctly different regimes of growth postulated by Galor (2005a, 2008a, 2011, 2012a) did not exist. His postulate ignores the data. There was no transition from stagnation to growth because there was no stagnation. During the alleged post-Malthusian regime, when the economic growth was supposed to start to follow a fast-increasing trajectory after the alleged epoch of stagnation, economic growth was diverted to a *slower* trajectory.

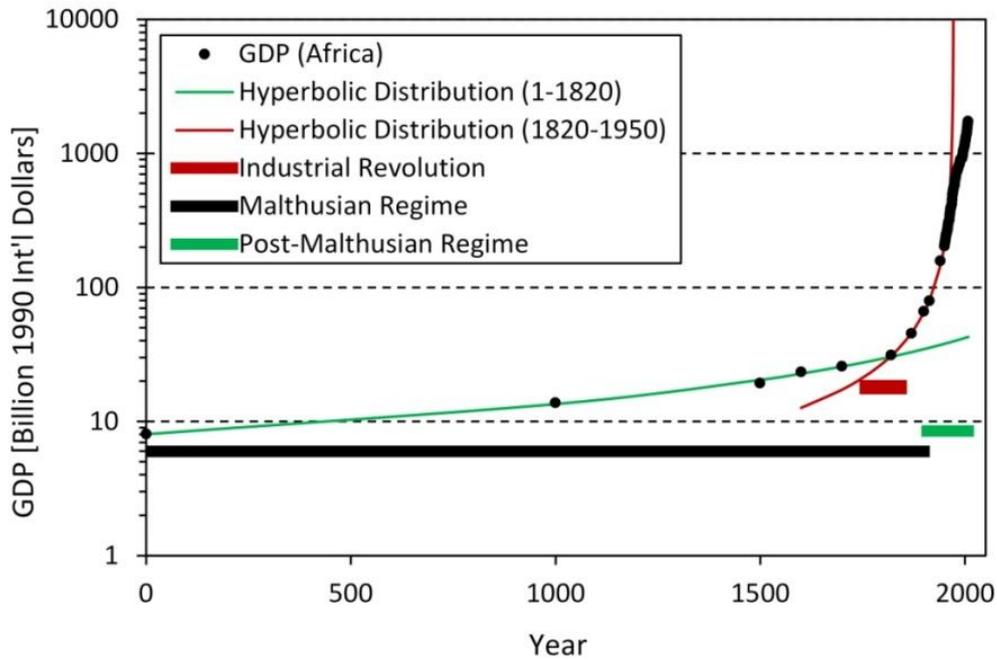

**Figure 17.** GDP data (Maddison, 2010) for Africa between AD 1 and 2008 compared with hyperbolic distributions. The two distinctly different regimes of growth postulated by Galor (2005a, 2008a, 2011, 2012a) did not exist. His postulate ignores the data. There was no stagnation and no transition to a faster growth at the end of the alleged regime of Malthusian stagnation. There was no escape from the Malthusian trap because there was no trap. During the alleged post-Malthusian regime, when the economic growth was supposed to start to follow a fast-increasing trajectory after the alleged epoch of stagnation, economic growth was diverted to a *slower* trajectory

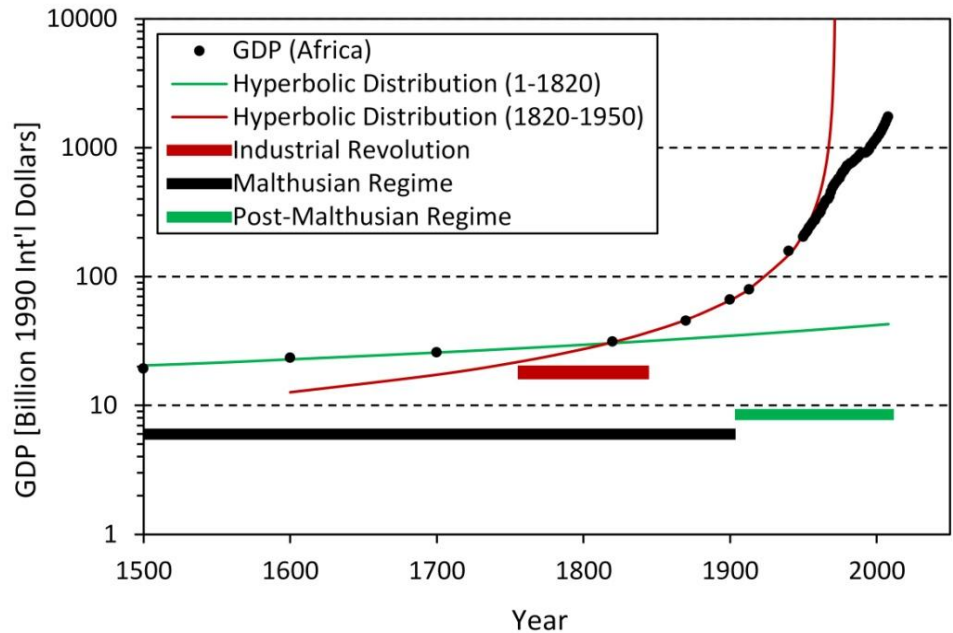

**Figure 18.** GDP data (Maddison, 2010) for Africa between AD 1500 and 2008 compared with hyperbolic distributions. The two distinctly different regimes of growth postulated by Galor (2005a, 2008a, 2011, 2012a) did not exist. His postulate ignores the data. The data are in clear contradiction of Galor's theory. There was no transition from stagnation to growth because there was no stagnation. During the alleged post-Malthusian regime, when the economic growth was supposed to start to follow a fast-increasing trajectory after the alleged epoch of stagnation, economic growth was diverted to a *slower* trajectory

**Latin America**

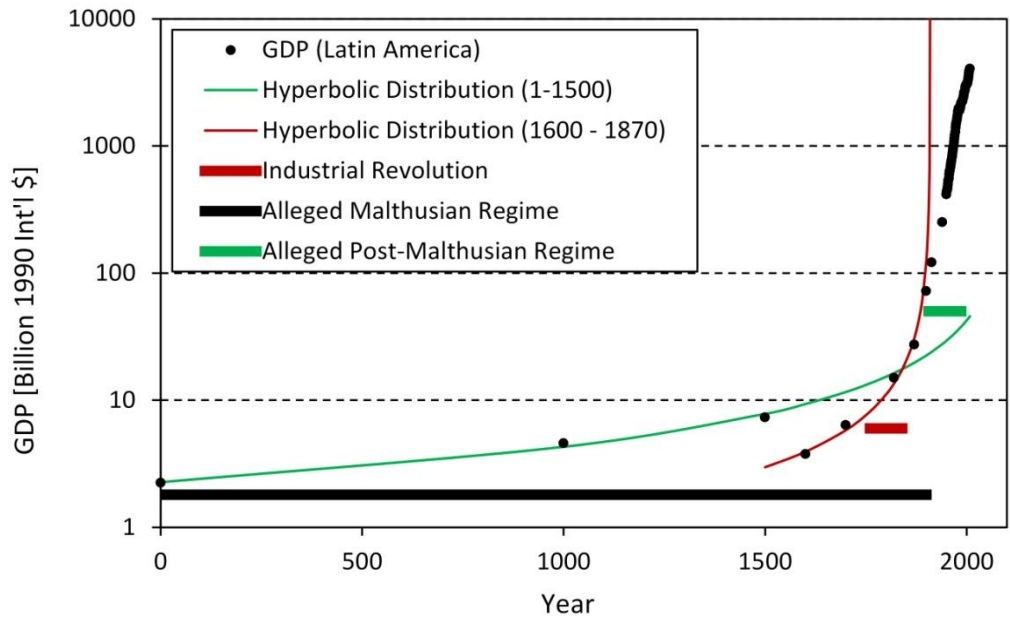

**Figure 19.** Economic growth in Latin America between AD 1 and 2008. Maddison's data (Maddison, 2010) are compared with hyperbolic distributions and with their unsubstantiated interpretations proposed by Galor (2005a, 2008a, 2011, 2012a). The data suggest two hyperbolic distributions, the pattern similar to the economic growth in Africa. The alleged transition from stagnation to growth never happened because the economic growth was not stagnant but hyperbolic. Around the time of the postulated by Galor "remarkable" escape from the alleged Malthusian trap (Galor, 2005a, p. 177) at the end of the alleged regime of stagnation, the economic growth started to be diverted to a *slower* trajectory. There was no escape from the Malthusian trap because there was no trap.

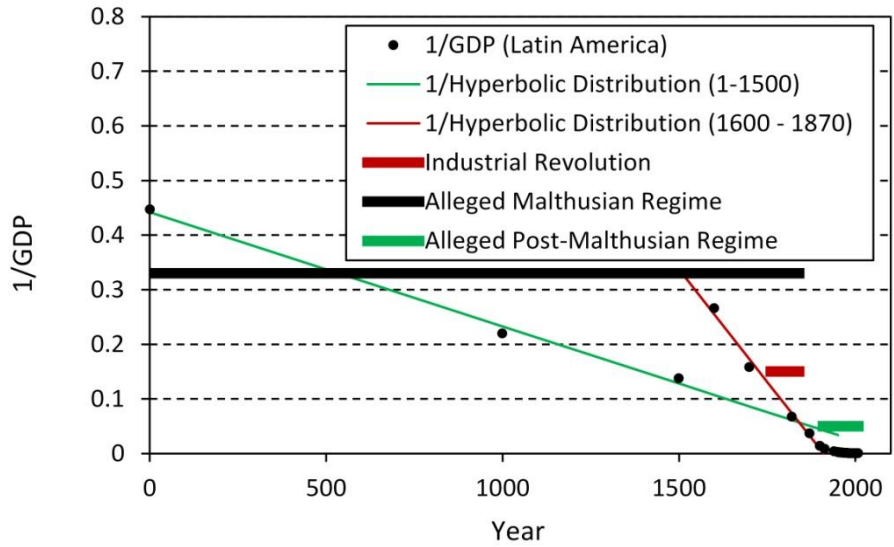

**Figure 20.** Reciprocal values of the GDP data, 1/GDP, for Latin America between AD 1 and 2008. Maddison's data (Maddison, 2010) are compared with hyperbolic distributions represented by the decreasing straight lines and with their unsubstantiated interpretations proposed by Galor (2005a, 2008a, 2011, 2012a). During the alleged regime of stagnation, the growth was hyperbolic. The data suggest two hyperbolic distributions, the pattern similar to the economic growth in Africa. The alleged transition from stagnation to growth around AD 1900 did not happened because there was no stagnation. Around the time of the alleged takeoff from stagnation to growth, the economic growth started to be diverted from the fast-increasing hyperbolic trajectory to a *slower* trajectory as indicated by the *upward* bending of the trajectory of the reciprocal values. There was no escape from the Malthusian trap because there was no trap. The transition from the slow to fast growth occurred around 300 years before the expected takeoff in AD 1900 and it was not a transition from stagnation to growth but from growth to growth. This feature, as well as the diversion to a slower trajectory at the time of the claimed takeoff around AD 1900, is not even noticed in the Unified Growth Theory.